# Microscopic and Spectroscopic Evidences for a Slater Metal-Insulator Transition in $Sr_2IrO_4$

Qing Li [1,‡], Guixin Cao [2,3,‡], Satoshi Okamoto [3,‡], Jieyu Yi [2,3], Wenzhi Lin [1], Brian C. Sales [3], Jiaqiang Yan [2,3], Ryotaro Arita [4,5], Jan Kunes [6], Anton V. Kozhevnikov [7], Adolfo G. Eguiluz [8], Masatoshi Imada [4,5], Zheng Gai [1], Minghu Pan [1,⋆], David G. Mandrus [2,3,⋆]

[1] *Center for Nanophase Materials Sciences, Oak Ridge National Laboratory, Oak Ridge, TN 37831, USA*

[2] *Department of Materials Science and Engineering, University of Tennessee, Knoxville, TN 37996, USA*

[3] *Materials Science and Technology Division, Oak Ridge National Laboratory, Oak Ridge, TN 37831, USA*

[4] *Department of Applied Physics, University of Tokyo, Hongo, Bunkyo-ku, Tokyo, 113-8656, Japan*

[5] *JST-CREST, Hongo, Bunkyo-ku, Tokyo, 113-8656, Japan*

[6] *Institute of Physics, Academy of Sciences of the Czech Republic, Cukrovarnicka´ 10, Praha 6, 162 53, Czech Republic*

[7] *Institute for Theoretical Physics, ETH Zurich, CH-8093 Zurich, Switzerland*

[8] *Department of Physics and Astronomy, University of Tennessee, Knoxville, TN 37996, USA*

‡ *These authors contributed equally to this work.*

⋆ e-mail: panm@ornl.gov and dmandrus@utk.edu

**Layered *5d* transition metal oxides (TMOs) have attracted significant interest in recent years because of the rich physical properties induced by the interplay between spin-orbit coupling, bandwidth and on-site Coulomb repulsion. In $Sr_2IrO_4$, this interplay opens a gap near the Fermi energy and stabilizes a $J_{eff}$=1/2 spin-orbital entangled insulating state at low temperatures. Whether this metal-insulating transition (MIT) is Mott-type (electronic-correlation driven) or Slater-type (magnetic-order driven) has been under intense debate. We address this issue via spatially resolved imaging and spectroscopic studies of the $Sr_2IrO_4$ surface using scanning tunneling microscopy/spectroscopy (STM/S). The STS results clearly illustrate the opening of the (~150-250 meV) insulating gap at low temperatures, in qualitative agreement with our density-functional theory (DFT) calculations. More importantly, the measured temperature dependence of the gap width coupled with our DFT+dynamical mean field theory (DMFT) results strongly support the Slater-type MIT scenario in $Sr_2IrO_4$. The STS data further reveal a pseudogap structure above the Neel temperature, presumably related to the presence of antiferromagnetic fluctuations.**

In *3d* TMOs, the localized *3d* states are responsible for the narrow *3d* bandwidth (*W*) and the accompanying strong on-site Coulomb repulsion (*U*). In contrast, in *5d* correlated electronic systems, the strong spin-orbit coupling (SOC) is comparable to *W* and *U*. The interplay between U, W and SOC gives rise to a broad spectrum of novel phenomena, *e. g.* the metal-insulator transition (MIT) may switch from Mott-type to Slater-type[1].

Recently, intensive studies have been carried out on the layered $5d$ TMO, $Sr_2IrO_4$, owing to the structural and electronic similarities to parent high-$T_c$ cuprates such as $La_2CuO_4$. It has been established that $Sr_2IrO_4$ exhibits insulating behavior[2,3], with predominant antiferromagnetic (AFM) order developing below 250 K[4]. However, debate continues as to whether $Sr_2IrO_4$ is more properly described as a Mott or a Slater insulator. Based on the calculated band structure, Kim *et al.*[5] proposed that $Sr_2IrO_4$ is a (canted) AF Mott insulator. In this scenario the interplay between SOC, $W$ and $U$ stabilizes a novel ground state electronic structure of $Sr_2IrO_4$, consisting a completely filled $J_{eff}=3/2$ bands and a narrow half-filled $J_{eff}=1/2$ band near the Fermi level ($E_F$). The $J_{eff}=1/2$ band is split into an upper Hubbard band (UHB) and lower Hubbard band (LHB) by a moderate Coulomb repulsion[6], exhibiting antiferromagnetic ordering of the effective $J_{eff}=1/2$ moments. In contrast, Arita *et al.*, have proposed a Slater mechanism for the gap formation in $Sr_2IrO_4$[7]. Such a Slater mechanism has been found in other $5d$ TMOs such as $NaOsO_3$[8].

In this article, we determine the atomically-resolved surface structure of $Sr_2IrO_4$ via scanning tunneling microscopy/spectroscopy (STM/S) and low-energy electron diffraction (LEED). Clear evidence is found to confirm the Slater-type nature of the insulating ground state in $Sr_2IrO_4$. This evidence includes the observation of an insulating gap in tunneling spectra, temperature-dependent spectroscopy measurements, and explicit comparison with DFT and DFT+DMFT calculations.

$Sr_2IrO_4$ crystallizes in the layered perovskite $K_2NiF_4$ structure and has octahedrally coordinated $Ir^+$ ions as shown in Fig. 1a[4]. Details of crystal growth and basic characterization are described in the Materials and Methods section of SOM. For the STM/S measurements, $Sr_2IrO_4$ single crystals were cleaved *in situ* under UHV environment at room temperature, producing a mirror-like surface. From crystal-chemical considerations and experience with other layered perovskites such as $Sr_2RuO_4$[9], the cleavage occurs between the Sr planes (see Fig.1a) and exposes the (001) surface with top Sr layer. After cleavage, the crystal was immediately inserted into the STM head at the base temperature of 4.2 K.

Figure 1b shows a 200 × 200 nm² topographic image of a cleaved $Sr_2IrO_4$ surface measured at 80 K in constant current mode. The height of a step is 12.8 Å, in good agreement with the single step height of the (001) terminated surface. By zooming into the area indicated by the white square box in Fig. 1b, a ''square-like'' lattice with an in-plane lattice constant of 3.8 Å can be barely observed (Fig. 1c), which reflects a termination of primary 1 × 1 structure truncated from a tetragonal bulk crystal (see Fig. 1a).

A typical LEED image at room temperature is shown in Fig. 1d. The unit cell derived from the LEED pattern appears to reflect a 1 × 1 surface for a tetragonal structure as observed by STM. In addition to the 1 × 1 pattern, additional spots are also clearly revealed, implying a √2 × √2 surface reconstruction. Such √2 × √2 surface reconstruction has been reported on other TMOs surfaces such as $Sr_2RuO_4$ and $Sr_3Ru_2O_7$[10,11] and can be attributed to $IrO_6$ octahedral rotation. The neutron diffraction measurements found the $IrO_6$ octahedra in $Sr_2IrO_4$ compound are indeed rotated around the crystallographic c axis by about 11°, thus reducing the space group symmetry from *I4/mmm* to *I4$_1$/acd* and generating an enlarged unit cell with dimensions √2a × 2c[4]. Such an octahedral rotation shown schematically in the inset of Fig. 1c, suggesting the origin of observed √2 × √2 surface reconstruction. Temperature-dependent LEED measurements (Fig. S1) show no structural change occurs on the (001) surface from room temperature to 150 K, therefore ruling out a structural transition on the surface.

In contrast to its *4d* counterpart $Sr_2RhO_4$ which is a 2D Fermi liquid, $Sr_2IrO_4$ is unexpectedly a canted antiferromagnetic insulator[4,12,13]. Magnetization measurements display weak ferromagnetism below 240 K with an easy axis along the *a*-axis[12]. Our electrical resistivity and magnetic measurements are in agreement with this conclusion.

STS probes the integrated surface density of electronic states (DOS) by measuring the tunneling conductance (the derivative of the tunneling current *dI/dV*). Averaged STS spectrum obtained at 80 K is shown in Fig. 1e; an insulating gap with width of approximately 0.20 eV is readily apparent. The blurring of gap structure is mainly due to the gap geometry in *k*-space. The gap symmetry can be investigated by tunneling into an arbitrary plane differing from (001). The DOS measured by an STS is simply an average over k-space. Most published STM experiments on cleaved oxides were carried out with the STM tip positioned perpendicular to the (001) surface. In this *c*-axis tunneling configuration, tunneling samples an angular average over the *ab*-plane density of states. Few attempt to do cross-sectional tunneling[14-17], *i.e.*, tunneling into an arbitrary plane parallel to the [001] direction, which is particularly attractive for high-$T_c$ cuprates. However, cross-sectional tunneling is a very challenging endeavor and even harder for atomic resolved cross-section imaging[18,19], primarily due to the difficulty in preparing a suitable cross-section for STM. Cleaving along other planes (instead of the (001) surface) is very unlikely because of the strong bonding between the Ir and O. However, occasionally some high-index facets could be found in our STM images after cleaving. STM/S measurements of the gap structure on these exposed facets proved key to understanding the microscopic origin of the insulating gap in $Sr_2IrO_4$.

Figure 2a shows a three dimensional, topographic STM image. Two crystallographic planes with different orientations are visible in this image. The surface in upper left corner is the normal (001) plane, which was observed most frequently in our STM imaging. After carefully checking the crystalline structure of $Sr_2IrO_4$, we can assign the facet in right corner to the (661) plane (see support materials Fig. S2 for details).

The atomic structures for the two orientated surfaces, (001) and (661), are shown in Fig. 2b, which exhibit a Sr-composed square and a mixed Sr-Ir pseudo-hexagonal lattice, respectively. Tunneling spectra at 4.2 K (Fig. 2c) exhibits different gap features at these two orientated surfaces. The gap on (001) surface is shallow with indistinct gap edges (blue curve in the upper panel). However, the gap observed on the (661) facet is sharper with well-defined edges (red curve in the bottom panel) compared to (001) surface. Such different gap features observed at two orientated surfaces suggest anisotropic gap geometry in $k$-space. In analogy to a $d$-wave superconductor, the insulating gap in $Sr_2IrO_4$ is angle-dependent. With different tunneling configurations, the DOS curves measured by an STM are a different average over $k$-space, therefore producing different gap features. By assuming certain gap landscapes for (001) and (661) surfaces (Fig. 2d), we are able to fit the gaps for each plane (see SOM for details). The gap geometries are proposed based on the symmetries of each plane, $e.g.$ (001) plane has four-fold symmetry, while the (661) plane is two-fold symmetric. From the fitting of the gap feature on (661) surface, we derive the gap size about 164 meV, indicating an insulating phase at low temperature. It is worth to note that the insulating gap of the bulk $Sr_2IrO_4$ is reported to be around 0.1 eV or less[20,21], smaller than our observed value on the surface. This fact is consistent with the conventional picture that electron-electron correlation effects should be stronger at the surface than that in the bulk as a result of the reduced atomic coordination[22]. We noticed two STM works recently posted in *Arxiv* by Nichols *et al.*[23] and Jixia Dai *et al.*[24], which reported the observation of an unusually large gap with the size up to 620 meV on $Sr_2IrO_4$ surface.

The atomically resolved STM image obtained at the (661) facet (Fig. 3a) shows a hexagonal lattice on this surface. The interatomic distance is about 3.7 Å, very close to the *a-axis* lattice constant of 3.78 Å. The exposed surface structure of (661) plane is composed of a mixture of Sr and Ir atoms located in same plane. The interatomic distance between Sr and Sr (green) is 3.78 Å, while the Ir-Ir and Sr-Ir distances are smaller (about 3.36 Å). Such discrepancy gives rise to a slight distortion of the hexagonal lattice. The distortion is clearly illustrated by the Fourier transform (FFT) image in Fig. 3b, which further confirms the measured surface is the (661) plane (Fig. 2b).

To obtain further insight into the variability of electronic structure on (661) surface, a spectroscopic survey was taken on the region marked by the blue square in Fig. 3a, which consists of differential tunneling conductance spectra (*dI/dV* vs. *V*) measured at a 40 × 40 grid in the given region. This 3D data set provides a detailed spatial map of the local density of states (LDOS) as a function of energy *E*. Most *dI/dV* spectra acquired exhibit similar gap features, implying such insulating gap is intrinsic and homogeneous.

Surprisingly, tunneling spectra at some locations (*e.g.* marked as a red star in Fig. 3a) have two prominent peaks at the gap edges (Fig. 3c), strongly resembling quasi-particle coherence peaks found in tunneling spectrum for superconductors[25,26]. Such peak-gap features can be fitted by BCS theory by considering anisotropic gap symmetry and a Dynes broadening factor $\Gamma$[27], which is convoluted with the Fermi function at 4.2 K. This resulted in the fitting parameters $\Gamma = 20$ meV and $\Delta = 160$ meV. The fitted curve (red line) is in good agreement with the STS data shown by the dotted curve. The quasi-particle lifetime can be estimated about $3 \times 10^{-14}$ sec from the inversion of Dynes broadening factor $\hbar/\Gamma$. The spatial mapping obtained by plotting the intensity of such peaks *vs.* spatial locations (Fig. 3d), reveals a coherence length $\xi \sim 5$ Å for quasi-particle excitations at 4.2 K. Short quasi-particle lifetime and small coherence length for low temperature insulating states in $Sr_2IrO_4$ suggest that some features of correlated $5d$ materials are similar to features found in BCS superconductors.

In order to gain further insight into the origin of the ground state in $Sr_2IrO_4$, we performed DFT calculations (details are given in the Method section). With a realistic value of the parameter $U = 2.5$ eV, the ground state is found to be a canted AF insulator, which is consistent with the experimental result. The resulting projected density of states (PDOS) is presented in Fig. 4a (the corresponding band structure is consistent with a previous study[20], as shown in the SOM, Fig. S3). The spectral intensity near $E_F$ is suppressed in a rather wide energy range of 200 meV. This feature is found to be consistent with the STS measurements. At reasonably high temperatures, the spectral shape is expected similar to the one shown in Fig. 4b, where the magnetic ordering is suppressed with finite $U$.

The distinction between the Slater transition and the Mott transition is rather subtle as, even in the Slater systems, local moments could exist above the magnetic transition temperature due to non-zero interactions. However, the Slater mechanism entails strong correlation between magnetic ordering and the opening of the insulating gap. In order to see such a correlation near $T_{MIT}$ more clearly, we extended our previous DFT-DMFT calculations for $Sr_2IrO_4$[7]. The resulting spectral functions are presented in Fig. 4c for various temperatures. While at high temperatures, we obtain metallic paramagnetic solutions, the system clearly undergoes a transition to the AF phase at the Néel temperature ($T_N$). In our calculation, the gap amplitude is increased as temperature is further lowered and the staggered magnetization is increased. Fig. 4d compares the staggered magnetization $M$ and the gap amplitude as a function of temperature. We note that we define the gap as half of the energy splitting between the upper Hubbard band (UHB) and lower Hubbard band (LHB), as the low-energy edge is hard to resolve. We emphasize that one can clearly see that the gap amplitude follows the staggered magnetic moment. As shown below, this behavior resembles our experimental measurements and strongly supports the Slater-type MIT in $Sr_2IrO_4$.

Figure 5a shows the temperature dependent STS spectra in the temperature interval 4.2 K-300 K. The evolution of the gap with temperature can provide critical information for the nature of the metal to insulator transition. It is noteworthy that a rigorous tip-conditioning procedure has been adopted in order to maintain a sharp and metallic tip during our experiments. Tip contamination affects the spectroscopic measurements and could lead to unusally large gap.

At each temperature the gap value is obtained by averaging spectra from an area of 100 nm × 100 nm by using a grid spectroscopic mode with 60 × 60 sampling pixels. To minimize the effect of sample inhomogeneity, such grid spectroscopic measurements at different temperatures were obtained at almost the same surface region by manually tracking the location by accounting for the thermal drift of the STM. The STS spectrum at room temperature displays no sign of an energy gap. An energy gap opens in the STS spectra below $T_N$ (Fig. 5a), indicating the establishment of an insulating state at the surface.

To elucidate the correlation of these changes, we have analyzed the STS spectra to extract information about the T dependence of the energy gap (Fig. 5b). With temperature increasing from 4 K, the gap value first increases from 150 meV to 250 meV at 100 K, followed by a continuous drop to zero at room temperature. The error in the estimated gap size originates from the uncertainty in determining the gap feature when the temperature is near $T_N$ (240 K). A BCS gap function (blue dashed line) is also plotted for comparison. We expect that the temperature dependence of a Slater insulating gap will behave in much the same way as a BCS gap[28]. As is clear from Fig. 5b, the temperature dependence of the insulating gap of $Sr_2IrO_4$ is consistent with the Slater scenario at least for temperatures between about 100 and 240 K.

The discrepancy between the experimental (Fig. 5b) and theoretical (Fig. 4d) temperature dependences of the energy gap can be understood by considering the temperature dependence of the magnetization of $Sr_2IrO_4$ with a magnetic field applied along the *a* or *c* directions[29,30]. According to a previous report[29], the easy axis is aligned in *a* direction and the canted AF order occurs around 240 K. In addition, the ac susceptibility starts to increase at around 50 K and reaches a peak at around 135 K, which is much lower than the magnetic transition temperature[29]. This peak reflects a metamagnetic transition associated with a strong enhanced electric permittivity peak around 100 K. The increase of the dielectric constant near 100 K (about 1 order of magnitude) increases the effective bias voltage between the STM tip and the sample, as well as the measured energy gap. The similarities in the temperature dependences of the insulating gap and the reported magnetization of $Sr_2IrO_4$[29], strongly imply a magnetic origin for the insulating state, consistent with a Slater type transition.

It is interesting to note, however, that in STS we observe "U-shaped" spectra above $T_N$. As the thermal broadening is estimated to be about 100 meV (4 $k_BT$) at room temperature, such U-shaped spectra indicate the existence of a pseudogap, which is absent in single-site DFT-DMFT analyses (Fig. 4c). However, the evidence of a pseudogap is consistent with the optical conductivity[3], which shows the closing of the gap without the Drude response, indicating a bad metal. It remains an open question whether or not non-local effects can reproduce such pseudogap[31-33] in $Sr_2IrO_4$ or whether additional degrees of freedom are necessary, such as electron-phonon coupling[34], in $Sr_2IrO_4$. STM/S technique is a surface sensitive probe. Near the surface, correlation effects are expected be stronger than in the bulk[35,36] because of the smaller coordination number and concomitant suppressed screening. Interestingly, in contrast to $Ca_{2-x}Sr_xRuO_4$[37], the MIT in $Sr_2IrO_4$ does not involve a structural transition and the closing of the gap coincides with the bulk magnetic transition. Thus, $Sr_2IrO_4$ may be less strongly-correlated in the bulk.

In summary, we studied the metal-insulator transition and low temperature insulating state in $Sr_2IrO_4$ using STM/S techniques and found a BCS-like gap feature with short quasi-particle lifetimes and a gap that opened smoothly as $Sr_2IrO_4$ was cooled below its Neel temperature. These results are most consistent with a Slater transition. Although the possibility of a Slater transition and the coexistence of Slater- and Mott-Hubbard-type behaviors have occasionally been discussed in the study of metal-insulator transitions in the iridates[38], our tunneling spectroscopic results, combined with our DFT+DMFT theoretical studies, provide unambiguous evidence that $Sr_2IrO_4$ is indeed on the Slater side. The spectra further support an evolution of a strongly correlated metal above the Neel temperature characterized by the pseudogap formation.

**Figure Captions**

**Figure 1 | Characterization of (001) cleaved $Sr_2IrO_4$ surface. a,** A ball model of the crystal structure of $Sr_2IrO_4$. The surface was created by cleavage between two SrO layers without breaking $IrO_6$ octahedra. **b,** A typical topographic image of the (001) surface. The setup conditions for imaging were a sample-bias voltage of 50 mV and a tunneling current of 0.1 nA. The right panel shows a cross-section along the line in the left panel. **c,** Atomically resolved image shows a square lattice of Sr atoms. Inset: Ball model of the surface structure with rotated octahedra (top view). **d,** A typical LEED image at the cleaved (001) surfaces of $Sr_2IrO_4$ measured at room temperature. LEED pattern shows $1 \times 1$ and $\sqrt{2} \times \sqrt{2}$ fractional spots (marked

by arrows). **e,** Averaged tunneling spectrum taken at 80 K, showing the opening insulating gap (averaged over an area of 100 nm × 100 nm by a grid spectroscopic mode with 20 × 20 sampling pixels). Data were taken with a sample bias voltage of –50 mV and a tunneling current of 0.1 nA. Bias-modulation amplitude was set to 3 mV$_{rms}$.

**Figure 2 | Crystallographic planes with different orientations. a,** A 3D landscape of scan area, showing two differently orientated surfaces after cleaving. The left facet is the (001) plane, while the right facet is determined as (661) plane. **b,** The ball models of surface atomic structures for (001) plane and (661) plane. The basic surface unit cells are marked by blue dashed lines. **c,** Local tunneling spectra were taken on the locations within different planes marked as stars in panel A, showing the variation of the insulating gap in different crystalline planes. The gap size can be estimated to be approximately 164 meV from the fitting of spectra. The fitting curves are shown as black lines in the same plots as the experimental STS data, based on the assumption of gap geometries in (001) and (611) planes. All spectral surveys were taken with a sample-bias voltage of 20 mV, a tunneling current of 0.1 nA, and bias modulation amplitude of 3 mV$_{rms.}$ **d,** The speculated gap geometries for facets (001) (upper panel) and (661) (bottom panel).

**Figure 3 | Atomic structure and BCS-like gap on (661) surface. a,** Atomically resolved STM image on (661) surface, showing a distorted hexagonal atomic structure. Scan size is 5 nm × 5 nm. The image was taken at the bias of 50 mV with 100 pA tunneling current. **b,** The FFT image displays a pseudo-hexagonal reciprocal lattice. **c,** Point tunneling spectrum taken at the location marked as star in panel (**a**) shows a well-defined gap feature and two quasi-particle coherence

peaks at the gap edge. Such gap and peaks can be well fitted by BCS theory with considering anisotropic gap in K space. The fitting details are described in support materials. The fitting parameters: gap $\Delta$=160 meV, Dynes broadening factor $\Gamma$=20 meV. **d,** A zoomed-in spectroscopic image with energy cut at peak position (160 meV), showing the quasi-particle coherence peaks decaying within a short spatial distance (about 5 Å). **e,** A series of spectra acquired along the yellow line in panel (**d**). The highlighted red curves have distinct coherence peaks at the edge of the gap while the other curves do not.

**Figure 4 | Density of states, elevated temperature calculations of $Sr_2IrO_4$. a,** Partial density of states of $Sr_2IrO_4$ projected onto Ir $d$ states with consideration of AF ordering. **b,** Same as A but without AF ordering. **c,** A series of PDOS projected onto $J_{eff}$=1/2 and 3/2 states of Ir with temperatures varying up to $T_N$. The gap value for temperature is estimated by the splitting between lower and upper Hubbard band as indicated. **d,** Staggered magnetization $M$ and the gap width as a function of temperature.

**Figure 5 | Temperature dependence of the insulating gap measured by STS. a,** Temperature dependence of scanning tunneling $dI/dV$ spectra measured at the surface of $Sr_2IrO_4$ (averaged over an area of 100 nm × 100 nm by a grid spectroscopic mode with 20 × 20 sampling pixels, sample bias $V$ = -50 mV, and feedback current $I$ = 100 pA). The spectra are shifted for clarity. The inset is the enlargement of the spectra near $E_F$ showing the opening energy gap (indicated by arrows). **b,** Temperature dependence of the energy gap determined by the STS spectra. The blue dashed curve is the temperature dependence of the energy gap for BCS theory for comparison.

**Acknowledgements**

This research was conducted (MP, QL, ZG) at the Center for Nanophase Materials Sciences, which is sponsored at Oak Ridge National Laboratory by the Scientific User Facilities Division, Office of Basic Energy Sciences, U.S. Department of Energy. Research was supported (GXC, SO, WL, JYY, BCS, JQY, DGM) by the U.S. Department of Energy, Basic Energy Sciences, Materials Sciences and Engineering Division. RA and MI were supported by MEXT Japan and Strategic Programs for Innovative Research (SPIRE), MEXT, and the Computational Materials Science Initiative (CMSI), Japan. JK was supported by the Grant No. 13-25251S of the Grant Agency of the Czech Republic. AVK acknowledges the computational resources of the CSCS and of the NCCS and the CNMS at ORNL, which are sponsored by the respective facilities divisions of the offices of ASCR and BES of the U.S. DOE. AGE was supported by NSF Grant No. OCI-0904972.

**Author contributions**

Q.L., W.L., M.P. performed STM measurements and analyzed STM data. G.X.C., J.Q.Y., B.C.S., D.G.M. grew and characterized the crystals. J.Y.Y., Z.G. performed the LEED measurements. S.O., R.A, J.K. A.V.K., A.G.E., M.I. carried out theoretical calculations. All authors discussed the results and M. P. wrote the paper with help from all authors. We thank Sergei V. Kalinin for careful reading.

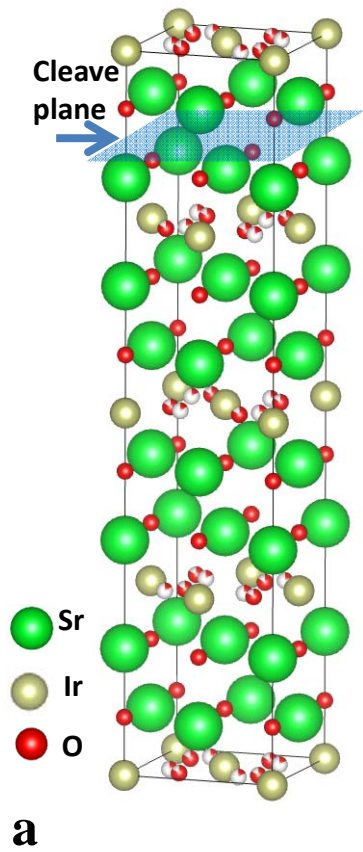
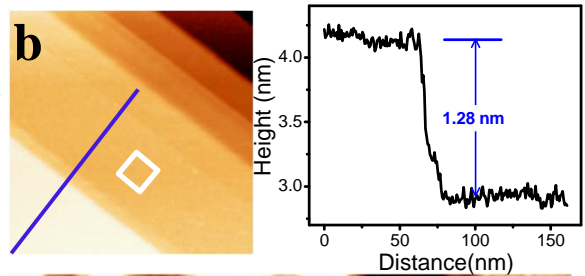
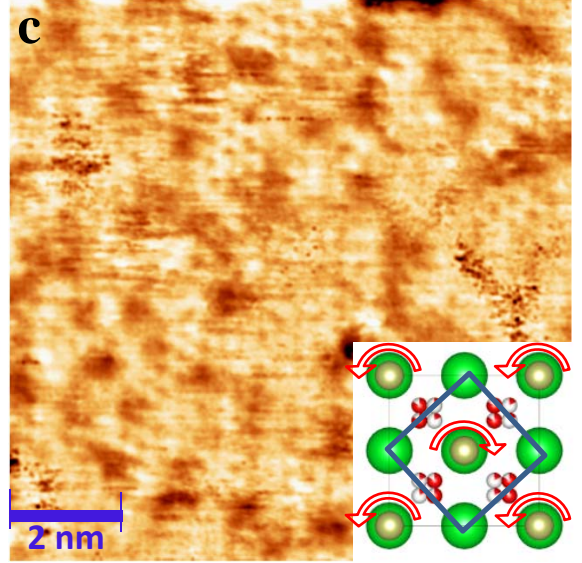
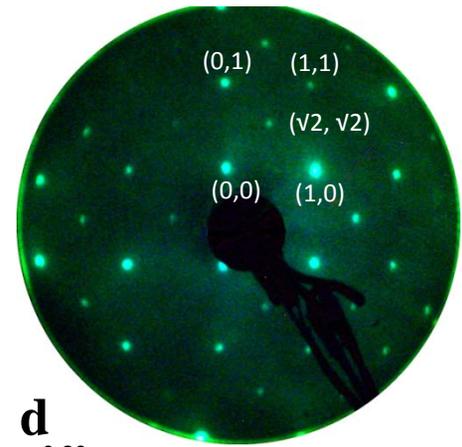
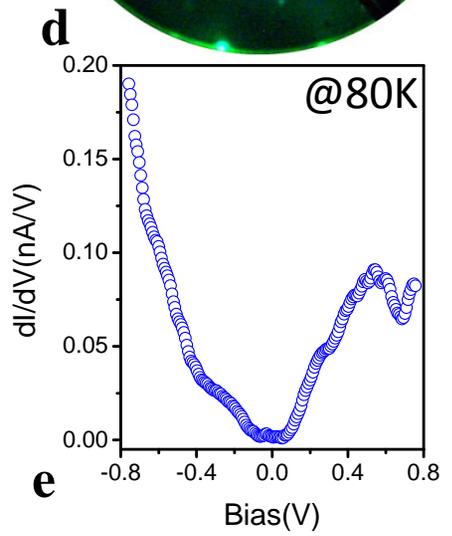

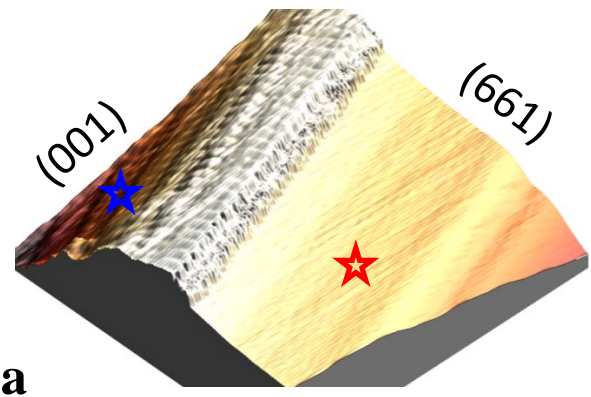
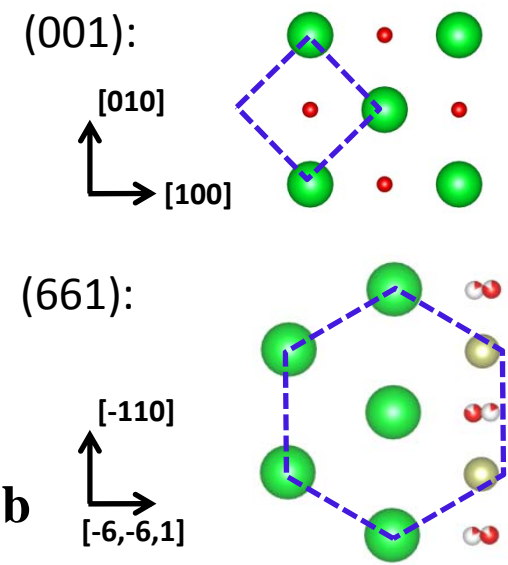
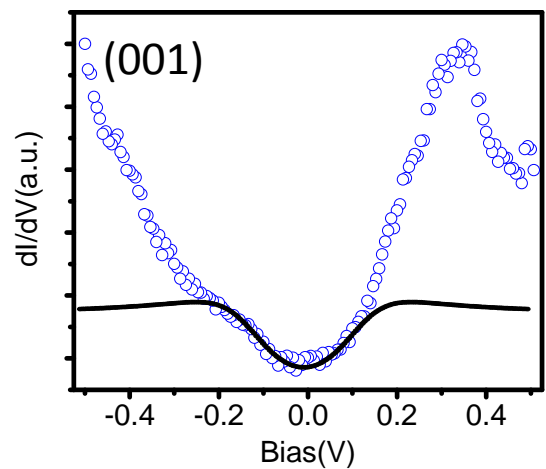
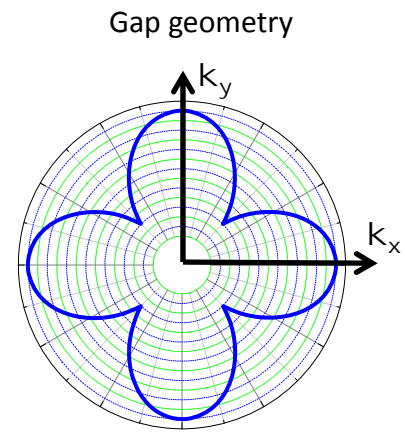
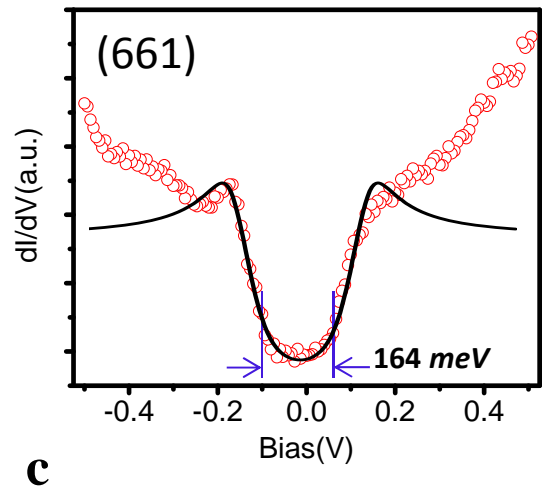
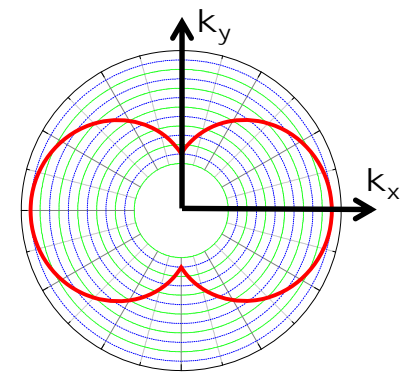

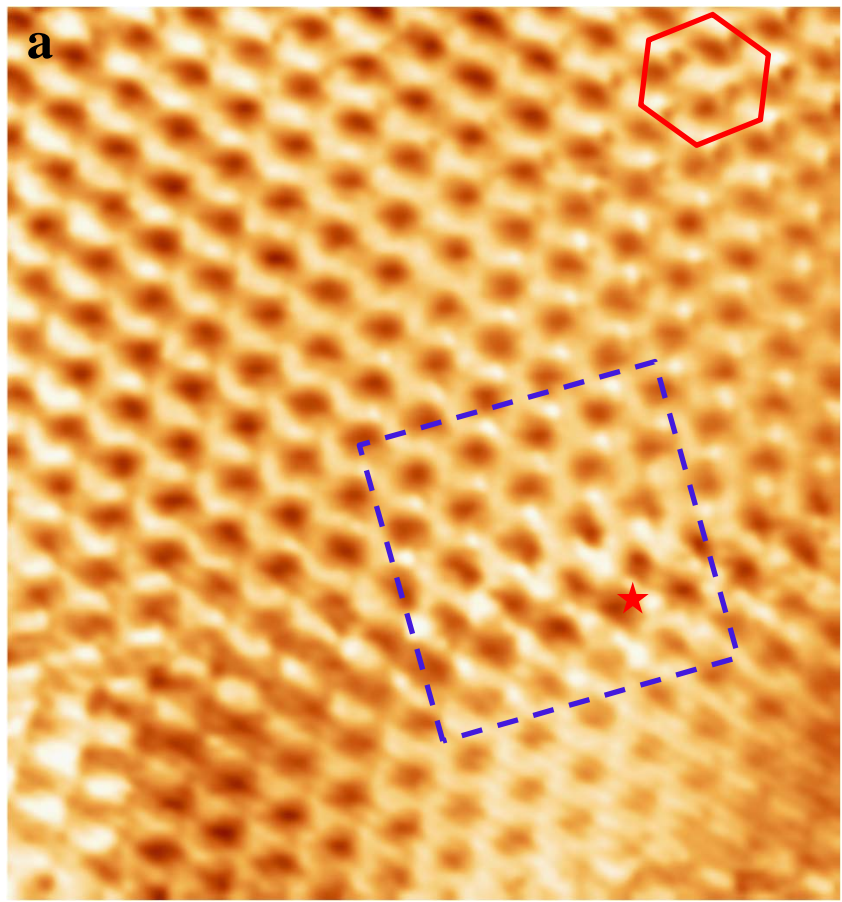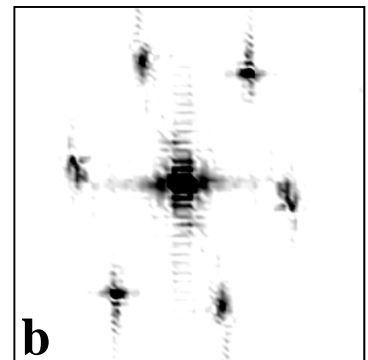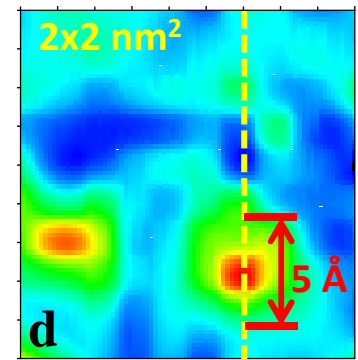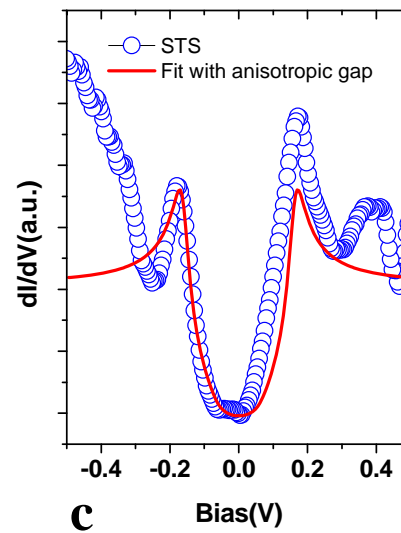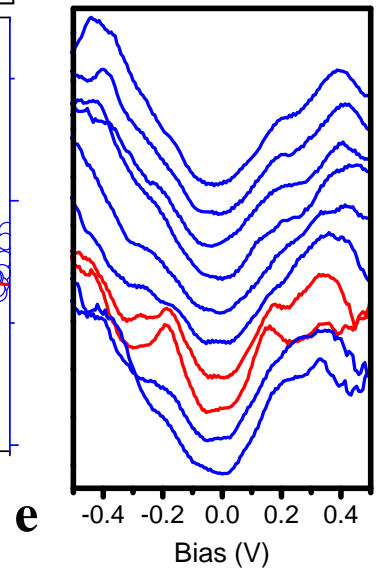

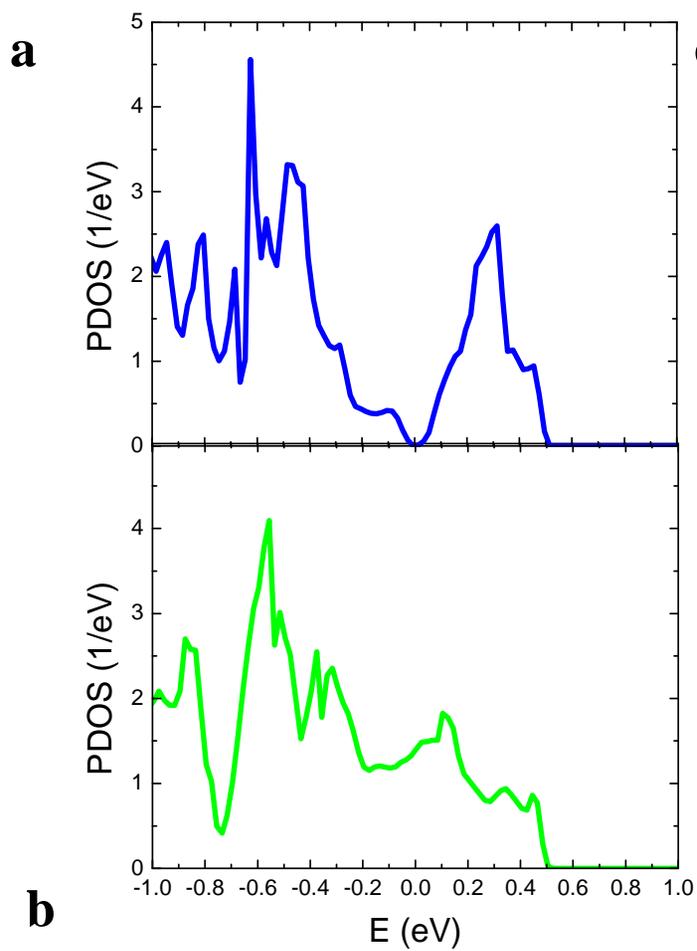
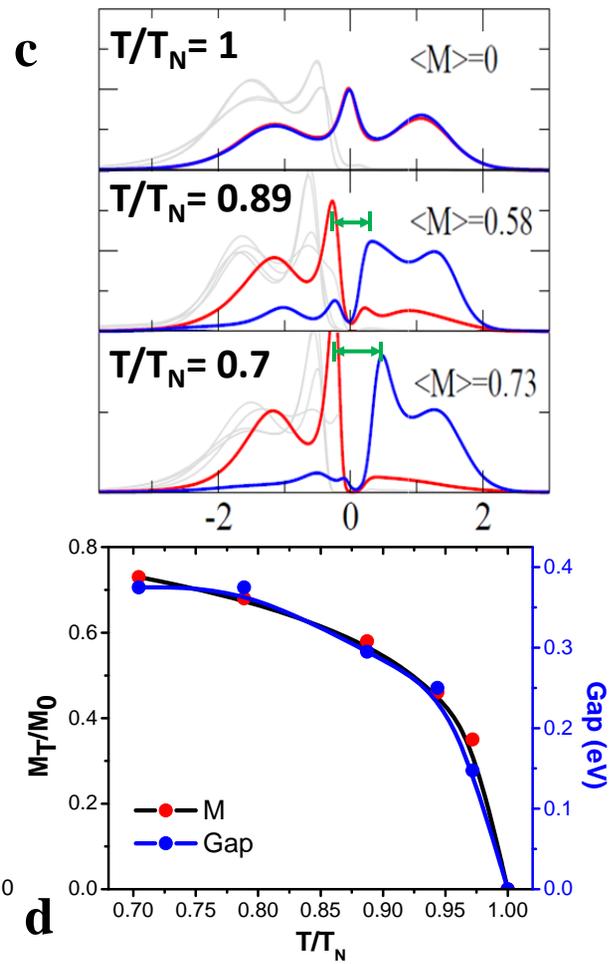

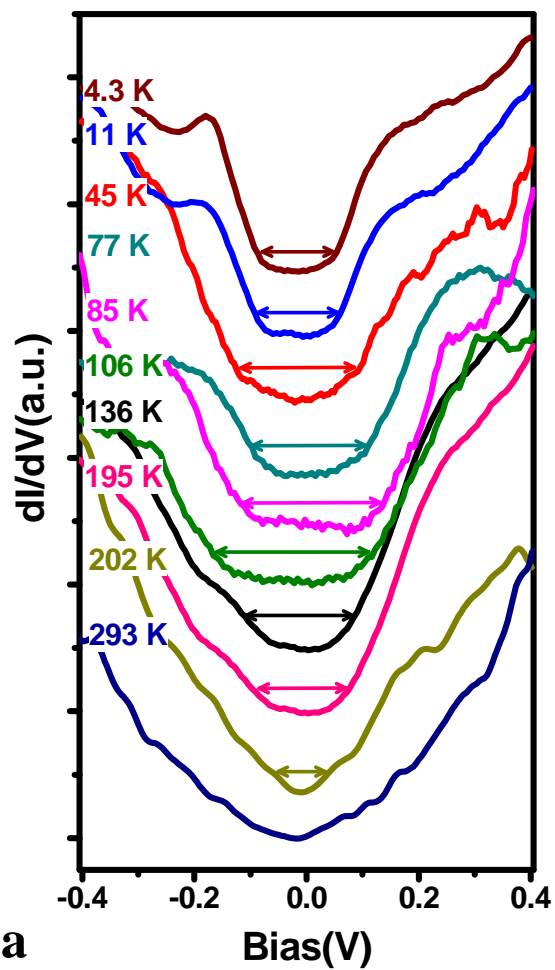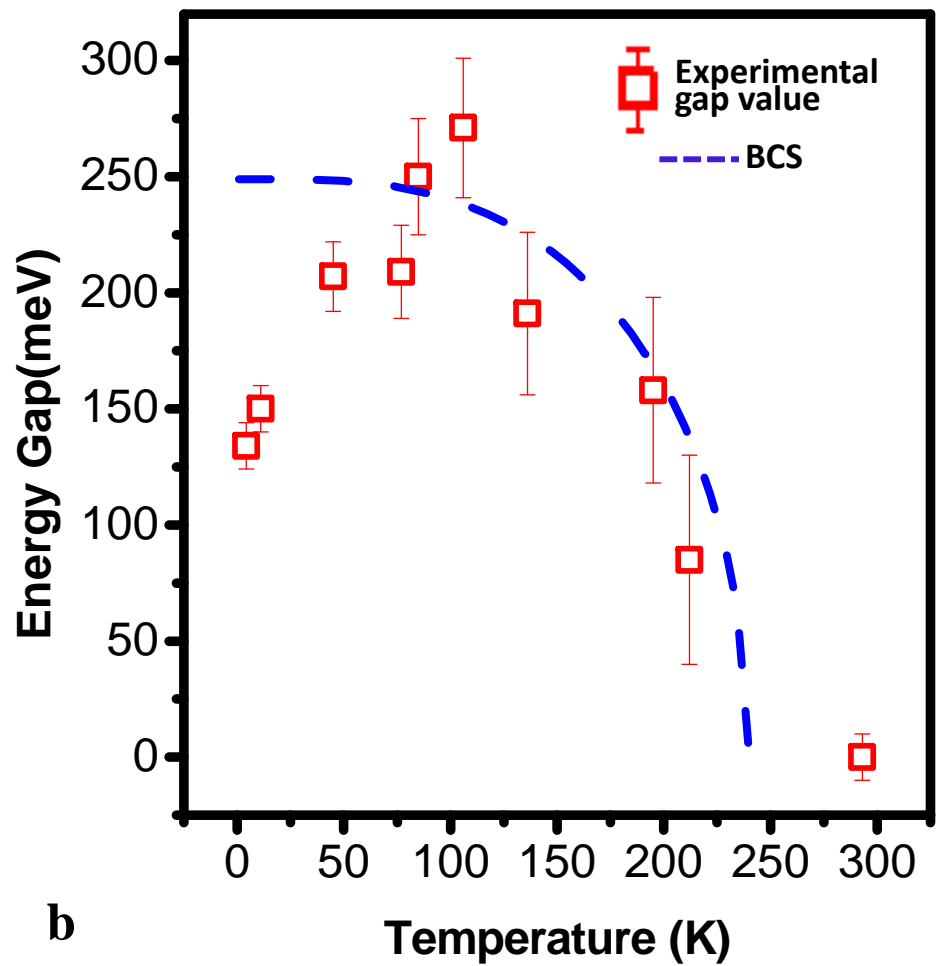